\documentclass[11pt]{article}
\usepackage{amsmath,amssymb,amsthm} 
\usepackage{bbm}
\usepackage{stmaryrd}
\usepackage{paralist}
\usepackage{hyperref}
\usepackage{color}

\setdefaultenum{(i)}{}{}{}
\usepackage[margin=1in]{geometry} 
\usepackage[pdftex]{graphicx}
\usepackage{subfigure}
\theoremstyle{plain}
 \numberwithin{mythm}{section}

\DeclareMathAlphabet\scr{U}{scr}{m}{n}
\SetMathAlphabet\scr{bold}{U}{scr}{b}{n}
  \DeclareFontFamily{U}{scr}{\skewchar\font'177}%
  \DeclareFontShape{U}{scr}{m}{n}{<-6>rsfs5<6-8>rsfs7<8->rsfs10}{}%
  \DeclareFontShape{U}{scr}{b}{n}{<-6>rsfs5<6-8>rsfs7<8->rsfs10}{}%

\numberwithin{equation}{section}

\renewenvironment{thebibliography}[1]{%
\begin{oldthebibliography}{#1}%
\setlength{\baselineskip}{.9em}
\linespread{.9}
\small
\setlength{\parskip}{0ex}%
\setlength{\itemsep}{.1em}%
}%
{%
\end{oldthebibliography}%
}

\DeclareMathAlphabet\scr{U}{scr}{m}{n}
\SetMathAlphabet\scr{bold}{U}{scr}{b}{n}
  \DeclareFontFamily{U}{scr}{\skewchar\font'177}%
  \DeclareFontShape{U}{scr}{m}{n}{<-6>rsfs5<6-8>rsfs7<8->rsfs10}{}%
  \DeclareFontShape{U}{scr}{b}{n}{<-6>rsfs5<6-8>rsfs7<8->rsfs10}{}%

\newcommand{\mal}{\stackrel{\mbox{\tiny$\bullet$}}{}}

\numberwithin{equation}{section}

\begin{document}

\title{\vspace{-1.5cm}Option Pricing and Hedging with Small Transaction Costs\footnote{We are grateful to Ale{\v{s}} {\v{C}}ern{\'y}, Christoph Czichowsky, Paolo Guasoni, Marcel Nutz, and Mete Soner for fruitful discussions. We also thank Ren Liu for his careful reading of the manuscript, and two anonymous referees for their pertinent remarks. Part of this work was completed while the second author was visiting Columbia University. He thanks Ioannis Karatzas and the university for their hospitality.}}
\author{
Jan Kallsen
\thanks{Christian-Albrechts-Universit\"at zu Kiel, Westring 383, D-24098 Kiel, Germany, email \texttt{kallsen@math.uni-kiel.de}.}
\and
Johannes Muhle-Karbe\thanks{ETH Z\"urich, Departement f\"ur Mathematik, R\"amistrasse 101, CH-8092, Z\"urich, Switzerland, and Swiss Finance Institute, email \texttt{johannes.muhle-karbe@math.ethz.ch}. Partially supported by the National Centre of Competence in Research Financial Valuation and Risk Management (NCCR FINRISK), Project D1 (Mathematical Methods in Financial Risk Management), of the Swiss National Science Foundation (SNF).}
}
\date{}

\maketitle

\begin{abstract}
An investor with constant absolute risk aversion trades a risky asset with general It\^o-dynamics, in the presence of small proportional transaction costs. In this setting, we formally derive a leading-order optimal trading policy and the associated welfare, expressed in terms of the local dynamics of the frictionless optimizer. By applying these results in the presence of a random endowment, we obtain asymptotic formulas for utility indifference prices and hedging strategies in the presence of small transaction costs.
\end{abstract}




\newpage

\section{Introduction}

The pricing and hedging of derivative securities is a central theme of Mathematical Finance. In complete markets, the risk incurred by selling any claim can be offset completely by dynamic trading in the underlying. Then, there is only one price compatible with the absence of arbitrage, namely the initial value of the replicating portfolio. This line of reasoning is torn to pieces by the presence of even the small bid-ask spreads present in the most liquid financial markets. Transaction costs make (super-)replication prohibitively expensive \cite{soner.al.95}, thereby calling for approaches that explicitly balance the gains and costs of trading. 

An economically appealing choice is the utility-indifference approach put forward by Hodges and Neuberger \cite{hodges.neuberger.89} as well as Davis, Panas, and Zariphopoulou \cite{davis.al.93}.\footnote{Cf.\ Leland \cite{leland.85} for an alternative approach, where trading only takes place at exogenous discrete times.} For an investor with given preference structure, the idea is to determine a ``fair'' price by matching the maximal expected utilities that can be attained with and without the claim. Both \cite{hodges.neuberger.89} and \cite{davis.al.93} focus on investors with constant absolute risk aversion for tractability. Nevertheless, the numerical computation of the solution turns out to be quite challenging, involving multidimensional nonlinear free boundary problems already for plain vanilla call options written on a single risky asset with constant investment opportunities.

 In reality, transaction costs are \emph{small}, having further declined substantially since stock market decimalization in 2001.\footnote{The CRSP database reports drops by almost one order of magnitude from 2001 to 2010.}  Therefore, asymptotic expansions for small spreads have been proposed to ``reveal the salient features of the problem while remaining a good approximation to the full but more complicated model'' \cite{whalley.wilmott.97}. For small costs, a formal asymptotic analysis of the model of Davis et al.\ \cite{davis.al.93} has been carried out by Whalley and Wilmott \cite{whalley.wilmott.97}.\footnote{See Bichuch \cite{bichuch.11} for a rigorous proof.}  A different limiting regime, where absolute risk version becomes large as the spread tends to zero, is studied by Barles and Soner \cite{barles.soner.98}. In both cases, the computation of indifference prices boils down to the solution of certain inhomogeneous Black-Scholes equations, whereas the corresponding hedging strategies are determined explicitly at the leading order. 

For small costs, the present study provides formal asymptotics for essentially general continuous asset price dynamics and arbitrary contingent claims. As in the extant literature, we also focus on investors with constant absolute risk aversion, for which the cash additivity of the corresponding exponential utility functions allows to handle the option position by a change of measure.\footnote{Similar but more involved asymptotics for the optimal policy and the associated welfare can also be obtained for general utilities and with intermediate consumption, see Soner and Touzi \cite{soner.touzi.11} as well as the forthcoming companion paper of the present study \cite{kallsen.muhlekarbe.12}. However, it is then no longer possible to deal with option positions by a change of measure as in Section \ref{sec:3}.} Both with and without an option position, the leading-order optimal trading strategies consist of keeping the number of risky shares in a time and state dependent no-trade region around their frictionless counterparts. The width of the latter is determined by the following tradeoff: large fluctuations of the frictionless optimizer call for a wide buffer in order to reduce trading costs. Conversely, wildly fluctuating asset prices cause the investor's positions to deviate substantially from the frictionless target, thereby necessitating closer tracking. Accordingly, the ratio of local fluctuations -- measured in terms of the local quadratic variation both for the frictionless optimizer and the risky asset -- is the crucial statistic determining the optimal trading boundaries in the presence of small transaction costs.  The corresponding welfare loss -- and in turn the indifference price adjustments -- turn out to be given by the squared width of the no-trade region, suitably averaged with respect to both time and states. 

The pricing implications of small transaction costs depend on the interplay between the frictionless pure investment and hedging strategies. If no trading takes place in the absence of an option position, then the costs incurred by hedging the claims necessitate a higher premium. This changes, however, if trades prescribed by the hedge partially offset the rebalancing of the investor's pure investment position. Then, maybe surprisingly, a smaller compensation may in fact be sufficient for the risk incurred by selling the claims if transaction costs are taken into consideration.

 The remainder of the article is organized as follows. The main results -- explicit formulas for the leading-order optimal policy and welfare -- are presented in Section~2. Subsequently, we discuss how they can be adapted to deal with utility-indifference pricing and hedging. The derivations of all results are collected in Appendix A. They are based on formal perturbation arguments, applied to the martingale optimality conditions for a frictionless ``shadow price'', which admits the same leading-order optimal strategy and utility as the original market with transaction costs. A rigorous verification theorem is a major challenge for future research.
 
\section{Optimal Investment}

Consider a market with two assets, a riskless one with price normalized to one and a risky one trading with proportional transaction costs. This means that one has to pay a higher ask price $(1+\varepsilon)S_t$ when purchasing the risky asset but only receives a lower bid price $(1-\varepsilon)S_t$ when selling it. Here, $\varepsilon>0$ is the relative bid-ask spread and the mid price $S_t$ follows a general, not necessarily Markovian, It\^o process:
\begin{equation}\label{eq:assetdynamics}
dS_t=b_t^S dt+\sqrt{c^S_t} dW_t,
\end{equation}
for a standard Brownian motion $W$. In this setting, an investor with exponential utility function $U(x)=-e^{-px}$, i.e., with constant absolute risk aversion $p>0$, trades to maximize the certainty equivalent $-\frac{1}{p} \log E[e^{-p X^\phi_T}]$ over all terminal wealths $X^\phi_T$ at time $T$ corresponding to self-financing trading strategies $\phi$.\footnote{The costs of setting up and liquidating the portfolio are only incurred once and are therefore of order $O(\varepsilon)$. Hence, they do not impact the investor's welfare at the leading order $\varepsilon^{2/3}$, and we disregard them throughout.} The optimal number of shares in the absence of frictions ($\varepsilon=0$) is denoted by $\varphi_t$; we assume it to be a sufficiently regular It\^o process with local quadratic variation $d\langle\varphi\rangle_t/dt$, which is satisfied in most applications. 

The dynamics \eqref{eq:assetdynamics} are formulated in \emph{discounted} terms. If the safe asset earns a constant interest rate $r>0$, one can reduce to this case by discounting, replacing risk aversion $p$ by $e^{rT}p$.

\subsection{Optimal Policy}\label{ss:main}

For small transaction costs $\varepsilon$, an approximately optimal\footnote{That is, this strategy matches the optimal certainty equivalent, at the leading order for small costs. See Appendix~\ref{sec:A2} for more details.} strategy $\varphi^\varepsilon_t$ is to engage in the minimal amount of trading necessary to keep the number of risky shares within the following buy and sell boundaries around the frictionless optimizer $\varphi_t$:
\begin{equation}\label{eq:main}
\Delta\varphi^{\pm}_t = \pm \left(\frac{3}{2p}\frac{d\langle\varphi \rangle_t}{d\langle S \rangle_t} \varepsilon S_t \right)^{1/3}.
\end{equation}
The random and time varying no-trade region $[\varphi_t+\Delta\varphi_t^-,\varphi_t+\Delta\varphi_t^+]$ is symmetric around the frictionless optimizer $\varphi_t$, and its halfwidth is given by the cubic root of three parts:
\begin{enumerate}
\item[i)] the constant $3/2p$ that only depends on the investor's risk aversion but not on the underlying probabilistic model.
\item[ii)] the observable (absolute) halfwidth $\varepsilon S_t$ of the bid-ask spread.
\item[iii)] the fluctuations of the frictionless optimizer, measured in terms of its local quadratic variation $d\langle \varphi\rangle_t$, normalized by the market's local fluctuations $d\langle S\rangle_t$. Tracking more wildly fluctuating strategies requires a wider buffer to reduce trading costs. Conversely, large fluctuations in the asset prices cause large fluctuations of the investor's risky position, thereby necessitating closer tracking to reduce losses due to displacement from the frictionless target position.
\end{enumerate}

Unless the planning horizon is postponed to infinity \cite{constantinides.86,davis.norman.90,shreve.soner.94}, transaction costs introduce horizon effects even with a constant investment opportunity set \cite{liu.loewenstein.02}. Nevertheless, the \emph{local} dynamics of the frictionless optimizer $\varphi_t$ alone always act as a sufficient statistic for the asymptotically optimal trading boundaries \eqref{eq:main} -- the investor does not hedge against the presence of a small constant friction. These optimal trading boundaries are ``myopic'' in the sense that they are of the same form as for the \emph{local} utility maximizers considered by Martin \cite{martin.12}. By definition, these also behave myopically in the absence of frictions, unlike the exponential investors considered here, whose optimal policies generally include an intertemporal hedging term reflecting future investment opportunities.  Somewhat surprisingly, the generally different frictionless strategies enter the leading order trading boundaries in the same way through their local fluctuations.

More general preferences and intermediate consumption are studied by Soner and Touzi \cite{soner.touzi.11} and in a forthcoming companion paper of the present study \cite{kallsen.muhlekarbe.12}.

\subsection{Welfare}

The utility associated to the above policy can also be quantified, thereby allowing to assess the welfare impact of transaction costs. To this end, let $\mathrm{CE}$ and $\mathrm{CE}^\varepsilon$ denote the certainty equivalents without and with transaction costs $\varepsilon$, respectively, i.e., the cash amounts that yield the same utility as trading optimally in the market. Then, for small transaction costs $\varepsilon$:
\begin{equation}\label{eq:main2}
\mathrm{CE}^\varepsilon \sim \mathrm{CE} -\frac{p}{2}E_Q\left[\int_0^T (\Delta\varphi^+_t)^2 d\langle S \rangle_t\right],
\end{equation}
and this leading-order optimal performance is achieved by the policy from Section \ref{ss:main}. As the trading boundaries $\Delta\varphi_t^+$ are proportional to $\varepsilon^{1/3}$ by \eqref{eq:main}, the leading-order loss due to transaction costs is therefore of order $O(\varepsilon^{2/3})$ as in the Black-Scholes model~\cite{shreve.soner.94}. It is given by an average of the squared halfwidth of the optimal no-trade region. The latter has to be computed with respect to a clock that runs at the speed $d\langle S \rangle_t =c^S_t dt$ of the market's local variance, i.e., losses due to transaction costs accrue more rapidly in times of frequent price moves. Moreover, this average has to be determined under the marginal pricing measure $Q$ associated to the frictionless utility maximization problem, i.e., the equivalent martingale measure that minimizes the entropy with respect to the physical probability. The leading-order effect of transaction costs can therefore be interpreted as the price of a path-dependent contingent claim, computed under the investor's marginal pricing measure. 

As first pointed out by Rogers \cite{rogers.04} (also compare Goodman and Ostrov \cite{goodman.ostrov.10}), the utility loss due to transaction costs is composed of two parts. On the one hand, there is the displacement loss due to following the strategy $\varphi_t^\varepsilon$ instead of the frictionless maximizer $\varphi_t$. In addition, there is the loss due to the costs directly incurred by trading. Rogers observed that, for small transaction costs, the leading orders of these two losses coincide for the optimal policy. For investors with constant absolute risk aversion, we complement this by the insight that two thirds of the leading-order welfare loss are incurred due to trading costs, whereas the remaining one third is caused by displacement. Surprisingly, this holds irrespective of the model for the risky asset and the investor's risk aversion.

\section{Indifference Pricing and Hedging}\label{sec:3}

Due to the cash additivity of the exponential utility function, the above results can be adapted to optimal investment in the presence of a random endowment, thereby leading to asymptotic formulas for utility-based prices and hedging strategies. 

Indeed, suppose that at time $t=0$, the investor sells a claim $H$ maturing at time $T$, for a premium $\pi(H)$. Then, her investment problem becomes
$$\textstyle{\sup_{\phi}} E\left[-e^{-p (X^\phi_T+\pi(H)-H)}\right]=e^{-p\pi(H)} E[e^{pH}]\textstyle{\sup_{\phi}} E_{P^H}\left[-e^{-p X^\phi_T}\right],$$
where $P^H$ is the equivalent probability with density $dP^{H}/dP=e^{pH}/E[e^{pH}]$. Hence, we are back in the above setting of pure investment; only the dynamics of the risky asset $S_t$ change when passing from the physical probability $P$ to $P^H$, and the frictionless optimizer changes accordingly. Then, with small transaction costs $\varepsilon$, the optimal policy in the presence of the random endowment $-H$ corresponds to the minimal amount of trading to keep the number of risky assets within the following buy and sell boundaries around the frictionless optimizer $\varphi_t^H$:\footnote{Note that the square-bracket processes of continuous processes are invariant under equivalent measure changes, i.e., it does not matter whether they are computed under $P$ or $P^H$.} 
\begin{equation}\label{boundH}
\Delta\varphi^{H,\pm}_t=\left(\frac{3}{2p}\frac{d\langle\varphi^H \rangle_t}{d\langle S \rangle_t}\varepsilon S_t\right)^{1/3}.
\end{equation}
Hence, at the leading order, the optimal investment strategy in the presence of a random endowment again prescribes the minimal amount of trading to remain within a buffer around its frictionless counterpart, whose width can be calculated from the local variations of the latter. Again by appealing to the results for the pure investment problem, the corresponding certainty equivalent is found to be
\begin{align}\label{eq:utilityH}
\mathrm{CE}^{\varepsilon, H} \sim \mathrm{CE}^H -\frac{p}{2}E_{Q^H}\left[\int_0^T (\Delta\varphi_t^{H,+})^2 d\langle S \rangle_t \right],
\end{align}
where $\mathrm{CE}^H$ and $Q^H$ denote the frictionless certainty equivalent and minimal entropy martingale measure in the presence of the claim $H$, respectively. With this result at hand, the corresponding \emph{utility indifference price} $\pi^\varepsilon(H)$ of Hodges and Neuberger \cite{hodges.neuberger.89} can be computed by matching \eqref{eq:utilityH} with the investor's certainty equivalent \eqref{eq:main2} in the absence of the claim. At the leading order, it turns out that the frictionless indifference price $\pi^0(H)$ -- which makes the frictionless certainty equivalents $\mathrm{CE}^H, \mathrm{CE}$ with and without the claim coincide -- has to be corrected by the difference between the effects of transaction costs with and without the claim: 
\begin{equation}\label{eq:idp}
\pi^\varepsilon(H) \sim \pi^0(H)+\frac{p}{2}\left(E_{Q^H}\left[\int_0^T (\Delta\varphi^{H,+}_t)^2 d\langle S \rangle_t\right]  - E_Q\left[\int_0^T (\Delta\varphi^+_t)^2 d\langle S \rangle_t \right]\right).
\end{equation}
As a consequence, the indifference price can be either higher or lower than its frictionless counterpart, depending on whether higher or lower trading costs are incurred due to the presence of the claim. 

\subsection{Complete Markets} 

Even in the absence of frictions, utility-based prices and hedging strategies are typically hard to compute unless the market is \emph{complete}, and we first focus on this special case in the sequel. Then, there is a \emph{unique} equivalent martingale measure $Q$, and frictionless indifference prices coincide with expectations under the latter:
$$\pi^0(H)=E_Q[H].$$
Moreover, any claim $H$ can then be hedged perfectly by a replicating strategy $\Delta_t^H$:\footnote{As this notation indicates, this is just the usual \emph{delta hedge} in a Markovian setting, i.e., the derivative of the option price with respect to the underlying.}
$$H=E_Q[H]+\int_0^T \Delta^H_t dS_t.$$
Consequently, the random endowment can simply be removed from the optimal investment problem:
$$\textstyle{\sup_\phi} E\left[-e^{-p(x+\int_0^T \phi_t dS_t+\pi(H)-H)}\right]=e^{-p(\pi(H)-E_Q[H])}\textstyle{\sup_\phi} E\left[-e^{-p(x+\int_0^T (\phi_t-\Delta^H_t) dS_t)}\right].$$
As a result, the optimal strategy for investors with constant absolute risk aversion is to hedge away the random endowment and, in addition to that, invest as in the pure investment problem:
\begin{equation}\label{eq:investH}
\varphi^H_t=\varphi_t+\Delta^H_t.
\end{equation}
The (monetary) boundaries of the corresponding leading-order optimal no-trade region for small transaction costs $\varepsilon$ can in turn be determined by inserting \eqref{eq:investH} into \eqref{boundH}:
\begin{equation}\label{eq:pmH}
\Delta\varphi^{\pm, H}_t S_t=\pm \left(\frac{3}{2p} \frac{d\langle \varphi+\Delta^H\rangle_t}{d\langle S \rangle_t}S_t^4\right)^{1/3} \varepsilon^{1/3}.
\end{equation}
To shed more light on this first-order correction due to the presence of small transaction costs, write\footnote{For $d\varphi_t=b^\varphi_t dt+\sqrt{c^\varphi_t} dW_t$, this is obtained by setting $\Gamma^\varphi_t=\sqrt{c^\varphi_t/c^S_t}$ and $a^\varphi_t=b^\varphi_t-b^S_t \sqrt{c^\varphi_t/c^S_t}$; the argument for $\Delta^H$ is analogous.}
\begin{equation}\label{eq:repgamma}
d\varphi_t= \Gamma^\varphi_t dS_t+a^\varphi_t dt,  \quad  d\Delta^H_t=\Gamma^H_t dS_t +a^H_t dt.
\end{equation}
The \emph{gammas} $\Gamma_t^\varphi$ and $\Gamma_t^H$ describe the sensitivities (of the diffusive parts) of the strategies $\varphi_t$ and $\Delta_t^H$ with respect to price moves.\footnote{In a Markovian setting, It\^o's formula shows that these processes indeed coincide with the usual notion of an option's ``gamma'', i.e., the second derivative of the option price with respect to the underlying.} With this notation, 
\begin{equation}\label{eq:gamma}
\frac{d\langle \varphi+\Delta^H\rangle_t}{d\langle S \rangle_t}S_t^4=(\Gamma_t^\varphi S_t^2+\Gamma_t^H S_t^2)^2.
\end{equation}
Hence, the width of the no-trade region \eqref{eq:pmH} is determined by the \emph{cash-gamma} of the investor's portfolio, that is, the sensitivity of the frictionless optimal risky position to changes in the risky asset. If shocks to the risky asset cause the frictionless position to move a lot, then the investor should keep a wider buffer around it to save transaction costs. For the Black-Scholes model, where the frictionless optimal risky position in the pure investment problem is constant, \eqref{eq:pmH} reduces to the formula derived by Whalley and Wilmott \cite[Section 4]{whalley.wilmott.97}.

Inserting \eqref{eq:pmH} into \eqref{eq:idp}, the corresponding utility indifference price for small transaction costs is found to be:
\begin{align}\label{eq:pricegeneral}
\pi^\varepsilon(H) &\sim E_Q[H]+ \left(\frac{9p}{32}\right)^{1/3} \varepsilon^{2/3} E_Q\left[\int_0^T \left[\left(\frac{d\langle \varphi+\Delta^H \rangle_t}{d\langle S \rangle_t}\right)^{2/3}-\left(\frac{d\langle \varphi \rangle_t}{d\langle S \rangle_t}\right)^{2/3}\right]S_t^{2/3} d\langle S \rangle_t\right] \notag\\
&=E_Q[H]+ \left(\frac{9p}{32}\right)^{1/3} \varepsilon^{2/3} E_Q\left[\int_0^T \left[|\Gamma^\varphi_t S_t^2+\Gamma^H_t S_t^2|^{4/3}-|\Gamma^\varphi_t S_t^2|^{4/3}\right]\frac{d\langle S \rangle_t}{S_t^2} \right].
\end{align}
The correction compared to the frictionless model is therefore -- up to a constant -- determined by the $Q$-expected time-average of the difference between (suitable powers of) the future cash-gamma of the investor's optimal position with and without the option, scaled by the infinitesimal variance of the relative returns. For the Black-Scholes model, one readily verifies that \eqref{eq:pricegeneral} can be rewritten in terms of the inhomogeneous Black-Scholes equation of Whalley and Wilmott \cite[Section 3.3]{whalley.wilmott.97}.

Representation \eqref{eq:pricegeneral} implies that the investor should charge a higher price than in the frictionless case if delta-hedging the option increases the sensitivity of her position with respect to price changes of the risky asset. Conversely, a lower premium is sufficient if delta hedging the claim reduces this sensitivity.  The impact of trade size and risk aversion depends on the relative importance of investment and hedging. Since the underlying frictionless market is complete, the perfect hedge $\Delta_t^H$ and in turn its gamma $\Gamma_t^H$ are independent of the investor's risk aversion, but scale linearly with the number $n$ of claims sold. Conversely, the optimal investment strategy $\varphi_t$ and its gamma $\Gamma_t^\varphi$ are independent of the option position sold, but scale linearly with the inverse of risk aversion. Consequently, the comparative statics of utility-based prices and hedges with transaction costs -- which depend on both quantities -- are ambiguous in general. However, they can be analyzed in more detail if either the option position or the pure investment dominates.

\subsubsection*{Marginal Investment}\label{marginalinvedtment}

First, we focus on the case where the investor's primary focus lies on the pricing and risk management of her option position. This regime applies if the cash-gamma $\Gamma_t^\varphi$ of the pure investment under consideration is negligible compared to its counterpart $\Gamma_t^{nH}$ for the option position $nH$. In particular, this occurs if the risky asset is assumed to be a martingale with vanishing risk premium, as in Hodges and Neuberger \cite{hodges.neuberger.89}, so that no investment is optimal without the option position, or in the asymptotic regime of Barles and Soner \cite{barles.soner.98}, where the option position increases as the spread becomes small.

If the contribution of the pure investment strategy is negligible, Formula \eqref{eq:pricegeneral} for the indifference price per claim reads as:
\begin{equation}\label{eq:smallphi}
\frac{\pi^\varepsilon(nH)}{n} \sim E_Q[H]+\left(\frac{9p n \varepsilon^2}{32}\right)^{1/3}  E_Q\left[\int_0^T |\Gamma^H_tS_t^2|^{4/3}\frac{d\langle S\rangle_t}{S_t^2}\right].
\end{equation}
For a marginal pure investment, small transaction costs therefore always lead to a positive price correction compared to the frictionless case. The interpretation is that the transaction costs incurred by carrying out the approximate hedge necessitate a higher premium. The size of this effect depends on the relative magnitudes of trade size, risk aversion, and transaction costs. Trade size $n$ and risk aversion $p$ both enter the leading order correction through their cubic roots, and therefore in an interchangeable manner.\footnote{Barles and Soner \cite{barles.soner.98} consider the case where the product $pn\varepsilon^2$ converges to a finite limit. In the Black-Scholes model, they characterize the limiting price per claim as the solution to an inhomogeneous Black-Scholes equation. However, this limit does not coincide with the right-hand side of \eqref{eq:smallphi}, whose derivation assumes only transaction costs to be small while risk aversion is fixed.}

If the contribution of the pure investment strategy is negligible, the optimal trading strategy in the presence of the claims can be directly interpreted as a utility-based hedge. In view of \eqref{eq:pmH}, the latter corresponds to keeping the number of risky shares within a no-trade region around the frictionless delta-hedge $\Delta^{nH}_t=n\Delta^H_t$; the maximal monetary deviations allowed are:
\begin{equation}\label{eq:hedge_complete_martingale}
\Delta\varphi^{nH,\pm}S_t=\pm \frac{n^{2/3} \varepsilon^{1/3}}{p^{1/3}}\left(\frac{3}{2} (\Gamma^H_t S_t^2)^2\right)^{1/3}.
\end{equation}
Higher risk aversion induces closer tracking of the frictionless target here, leading to more trading and in turn higher prices (cf.\ Formula \eqref{eq:smallphi}). 

\paragraph{Semi-Static Delta-Gamma Hedging}

``Delta-gamma hedging'' is often advocated in order to reduce the impact of transaction costs, cf., e.g., \cite[p.\ 129]{bjoerk.03}. The above results allow to relate this idea to the semi-static hedging of a claim $H$, by dynamic trading in the underlying risky asset and a static position $n'$ in some other claim $H'$ set up at time zero. In the frictionless case, the choice of $n'$ does not matter, since any such position can be offset by delta-hedging with the underlying. With transaction costs, this no longer remains true. Suppose the risky asset is a martingale ($P=Q$) and $H$, $H'$ are traded at their frictionless prices $E_Q[H], E_Q[H']$ with some transaction costs of order $O(\varepsilon)$. Then, \eqref{eq:smallphi} applied to the claim $H-n' H'$ shows that the leading-order frictional certainty equivalent of selling one unit of $H$ -- and hedging it with a static position of $n'$ units of $H'$ and optimal dynamic trading in the underlying -- is given by:
$$ -\left(\frac{9p}{32}\right)^{1/3} \varepsilon^{2/3} E_{Q}\left[\int_0^T \left|\Gamma^H_t S_t^2 - n' \Gamma^{H'}_t S_t^2 \right|^{4/3} \frac{d\langle S \rangle_t}{S_t^2} \right].$$
Maximizing this certainty equivalent in $n'$ therefore amounts to minimizing a suitable average of the future (cash) gamma of the total option position $H-n' H'$. 

If other options are only used once for hedging, the total position should thus not be made gamma-neutral at any one point in time. The exception is when both option gammas are approximately constant over the horizon under consideration; then, it is optimal to make the total option position gamma-neutral. This situation occurs if the static option position is only held briefly, and before maturity of either claim. Hence, it seems reasonable to conjecture that one should trade to remain close to a delta-gamma neutral position if hedging dynamically with both the underlying and an option.\footnote{Compare \cite{goodman.ostrov.11} for related results in a pure investment problem with a stock and an option.}

\subsubsection*{Marginal Option Position}

Now, let us turn to the converse situation of a  \emph{marginal option position}, i.e., the sale of a \emph{small} number $n$ of claims $H$. For incomplete markets without frictions, the limiting price for $n\to 0$, called \emph{marginal utility-based price}, is a linear pricing rule, namely the expectation under the frictionless minimal entropy martingale measure, independent of both trade size and risk aversion. The leading-order correction for small $n$ is linear both in the trade size $n$ and in the investor's risk aversion $p$ \cite{mania.schweizer.05, becherer.06,kallsen.rheinlaender.11}. Hence, both quantities are interchangeable also in this setting: doubling risk aversion has the same effect as doubling the trade size.

Let us now derive corresponding results for the incompleteness caused by imposing small transaction costs in an otherwise complete market. Then, Taylor expanding \eqref{eq:pricegeneral}  for small $n$ yields:\footnote{Here and in the sequel, we report the leading-order terms for small transaction costs $\varepsilon$ \emph{and} small trade size $n$.}
\begin{align*}
\frac{\pi^\varepsilon(nH)}{n} \sim E_Q[H]&+ \left(\frac{9p}{32}\right)^{1/3} \varepsilon^{2/3} E_Q\left[\int_0^T\frac{4}{3} |\Gamma^\varphi_t S_t^2|^{4/3} \frac{\Gamma^H_t}{\Gamma^\varphi_t}\frac{d\langle S \rangle_t}{S_t^2}\right]
 \label{eq:pricesmalln}\\
&+n  \left(\frac{9p}{32}\right)^{1/3} \varepsilon^{2/3} E_Q\left[\int_0^T\frac{2}{9} |\Gamma^\varphi_t S_t^2|^{4/3} \left(\frac{\Gamma^H_t}{\Gamma^\varphi_t}\right)^2\frac{d\langle S \rangle_t}{S_t^2}\right].\notag
\end{align*}
Recall that the optimal frictionless strategy $\varphi_t$ and its gamma $\Gamma^\varphi_t$ are independent of the trade size $n$ but scale linearly with the inverse of risk aversion $p$. Hence, the frictionless scalings are robust with respect to small transaction costs: The limiting price per claim for $n \to 0$ is also linear in the claim, and independent of trade size and risk aversion. Moreover, these quantities both enter linearly, and therefore in an interchangeable manner, in the leading order correction term for small trade sizes. 

For a small option position, the sign of the prize correction compared to the complete frictionless market depends on the interplay between the pure investment strategy and the hedge for the claim. The pure investment strategy is typically negatively correlated with price shocks, $\Gamma_t^\varphi<0$ (e.g., in the Black-Scholes model). Then, the difference between the frictionless price and the limiting price with transaction costs is determined by the sign of the option's gamma. If the latter is positive as for European call or put options, then the marginal utility-based price taking into account transaction costs is smaller than its frictionless counterpart. This is because utility-based investment strategies are typically of contrarian type, i.e., decreasing when the risky price rises, whereas the delta-hedge of a European call or put is increasing with the value of its underlying. Consequently, hedging the claim allows the investor to save transaction costs, so that she is willing to sell the claim for a smaller premium.  This rationale, however, is only applicable if the fluctuations of the hedging position are small enough to be absorbed by the investor's other investments. In particular, the price adjustment is always positive for a marginal pure investment.

Let us also consider how the investor's trading strategy changes in the presence of a small number of claims. Taylor expanding \eqref{eq:pmH} for small $n$ shows that it is optimal to refrain from trading as long as the risky position remains within a bandwidth of
$$\Delta\varphi^{nH,\pm}_t S_t \sim  \pm \Delta\varphi^\pm_t S_t \left(1+ \frac{2}{3}n \frac{\Gamma^H_t}{\Gamma^\varphi_t}\right)$$
around the frictionless optimal position $(\varphi_t+n \Delta^H_t)S_t$. The interpretation for the ratio of gammas is the same as for the corresponding utility-based prices above: if the trades prescribed by the delta hedge partially offset moves of the pure investment strategy, then the resulting reduced sensitivity to price shocks allows the investor to use a smaller no-trade region than in the absence of the claims. As for risk aversion, note that whereas the investor's pure investment and the corresponding trading boundaries $\Delta\varphi_t^{\pm}$ in the absence of the claim scale with her risk tolerance $1/p$, the adjustment due to the presence of the claims does not. For small option positions, it is linear in trade size but independent of risk aversion as is true for the frictionless hedge.

\subsection{Incomplete Markets}

In incomplete markets, simple formulas for indifference prices and hedging strategies can typically only be obtained in the limit for a small number of claims, even in the absence of frictions. If the trade size $n$ is small, \cite{mania.schweizer.05, becherer.06, kallsen.rheinlaender.11} show that the optimal strategy $\varphi_t$ for the pure investment problem should be complemented be $n\xi_t$, where $\xi_t$ is the mean-variance optimal hedge for the claim, determined under the marginal pricing measure $Q$, i.e.,
$$\xi_t=\frac{d\langle V,S \rangle_t}{d\langle S \rangle_t},$$
where $V_t$ denotes the $Q$-martingale generated by the payoff $H$. As a consequence, the leading-order adjustment of the portfolio due to the presence of the claim is linear in trade size and independent of risk aversion, as in the complete case discussed above. The corresponding indifference price per claim converges to the expectation under the marginal pricing measure, which is again independent of trade size and risk aversion. The leading-order adjustment for larger trade sizes is given by the $pn/2$-fold of the minimal $Q$-expected squared hedging error, i.e.,
\begin{equation}\label{eq:hedgeingerror}
\frac{\pi^0(nH)}{n} \sim E_Q[H]+\frac{pn}{2} E_Q\left[\left(H-E_Q[H]-\int_0^T \xi_t dS_t\right)^2\right].
\end{equation}
As a result, it is linear both in trade size and risk aversion.  In this setting of a small option position held in a potentially incomplete frictionless market, we now discuss the implications of small transaction costs.

\subsubsection*{Negligible Risk Premium}

Let us first consider the case where the risky asset is a martingale under the physical probability. Then, no trading is optimal for the pure investment problem, $\varphi_t=0$, and the minimal entropy martingale measure coincides with the physical probability, $Q=P$. As a consequence, the monetary trading boundaries  \eqref{boundH} around the frictionless strategy $n\xi_t$ are given by:
$$\Delta\varphi^{nH,\pm}_t S_t \sim \pm \frac{n^{2/3}\varepsilon^{1/3}}{p^{1/3}} \left(\frac{3}{2} \frac{d\langle \xi\rangle_t}{d\langle S\rangle_t} S_t^4\right)^{1/3}.$$
In view of \eqref{eq:gamma}, this is the same formula as in the complete case \eqref{eq:hedge_complete_martingale}, with the perfect hedge replaced by the mean-variance optimal one. In particular, the scalings in trade size and risk aversion are robust to incompleteness in the frictionless market, as long as the option position is small.

To determine the corresponding leading order price correction, insert the above trading boundaries into \eqref{eq:idp} and note that $Q=P$, $\Delta\varphi_t^+=0$, as well as $dQ^{nH}/dP=1+O(n)$. As a consequence:
$$\frac{\pi^\varepsilon(nH)}{n} \sim E[H]+ \frac{pn}{2} E\left[\left(H-E[H]-\int_0^T \xi_t dS_t\right)^2\right] +\left(\frac{9pn \varepsilon^2}{32}\right)^{1/3} E\left[\int_0^T \left(\frac{d\langle \xi\rangle_t}{d\langle S\rangle_t} S_t^4\right)^{2/3} \frac{d\langle S\rangle_t}{S_t^2}\right].$$
The second term is the correction due to transaction costs, which once more parallels the complete case \eqref{eq:smallphi}, with the mean-variance optimal hedge again replacing the replicating strategy. The first term is the correction \eqref{eq:hedgeingerror} due to the incompleteness of the frictionless market, which is proportional to the minimal squared hedging error and hence vanishes in the complete case. At the leading order, the two price corrections therefore separate; their relative sizes  are determined by the magnitude of risk aversion $p$ times trade size $n$, compared to the spread $\varepsilon$.

\subsubsection*{Nontrivial Risk Premium}

If the pure investment strategy $\varphi_t$ is not negligible, the trading boundaries \eqref{boundH} around the frictionless strategy $\varphi_t+n\xi_t$ are given by
$$\Delta\varphi^{nH,\pm}_t \sim \Delta\varphi_t^{\pm}\left(1+\frac{2n}{3}\frac{ d\langle \varphi, \xi \rangle_t }{d\langle \varphi \rangle_t}\right).$$
In the small claim limit, the interpretations from the complete case are therefore robust as well: If shocks to the frictionless investment and hedging strategies are negatively correlated, one should keep a smaller buffer with the claim, and conversely for the case of a positive correlation. 

Concerning the pricing implications of small transaction costs added to incomplete frictionless markets, the situation is somewhat more involved. The reason is that the presence of the claim changes the impact of the transaction costs in two different ways. On the one hand, it affects the trading strategy that is used: the investor passes from staying within $\Delta\varphi^{\pm}_t$ around the pure investment strategy $\varphi_t$, to keeping within $\Delta\varphi_t^{nH,\pm}$ around $\varphi_t^{nH}=\varphi_t+n\xi_t+O(n^2)$. On the other hand, even after hedging the claim, the latter still induces some unspanned risk in incomplete markets, and therefore affects the investor's marginal evaluation rule. That is, the marginal pricing measure changes from $Q$, with density proportional to the marginal utility $U'(\int_0^T \varphi_t dS_t)$ associated to the pure investment strategy, to $Q^{nH}$, with density proportional to the marginal utility augmented by the $n$ claims, $U'(\int_0^T \varphi_t^H dS_t-nH)$  (compare \cite[Theorem 1.1]{owen.02}). By Formula \eqref{eq:idp} and Taylor expansion for small $n$, the leading-order price impact of small transaction costs is then found to be given by:
$$\frac{p}{2}\left(E_Q\left[\int_0^T \frac{4n}{3}(\Delta\varphi^+_t)^2\frac{d\langle \varphi,\xi\rangle_t}{d\langle \varphi \rangle_t}d\langle S\rangle_t \right]+E\left[\left(\frac{dQ^{nH}}{dP}-\frac{dQ}{dP}\right)\int_0^T (\Delta\varphi^+_t)^2 d\langle S\rangle_t\right] \right). $$
The first term is due to changing the trading strategy; it is already visible for complete frictionless markets. The second term reflects the change of the marginal pricing measure due to the presence of the claim, which does not take place in  complete frictionless markets with a unique equivalent martingale measure. As for the hedging strategy above, the sign of the first term depends on the correlation of shocks to investment and hedging strategies. To examine the sign of the second term, notice that $\varphi_t^H=\varphi_t+n\xi_t+O(n^2)$, Taylor expansion, and the $Q$-martingale property of the wealth process $\int_0^T \xi_t dS_t$ yield
$$\frac{dQ^{nH}}{dP}=\frac{e^{-p(\int_0^T(\varphi_t+n\xi_t) dS_t +O(n^2)-nH)}}{E[e^{-p(\int_0^T(\varphi_t+n\xi_t)dS_t +O(n^2)-nH)}]}=\frac{dQ}{dP}\left(1+np\left( H-E_Q[H]-\int_0^T \xi_t dS_t\right)\right)+O(n^2).$$
As a result, the second term in the price impact of small transaction costs is given by the covariance between the shortfall of the frictionless utility-based hedge and the cumulated transaction costs effect, measured by the average squared width of the no-trade region:
$$E\left[\left(\frac{dQ^{nH}}{dP}-\frac{dQ}{dP}\right)\int_0^T (\Delta\varphi^+_t)^2 d\langle S \rangle_t\right] \sim np E_Q\left[\left( H-E_Q[H]-\int_0^T \xi_t dS_t\right)\int_0^T (\Delta\varphi^+_t)^2 d\langle S \rangle_t\right].$$
Hence, incompleteness of the frictionless market increases the impact of transaction costs, if these tend to accrue more rapidly when the imperfect utility-based hedge also does badly, i.e., when the different sources of incompleteness tend to cluster. In contrast, the premium for the option is decreased if the two risks are negatively correlated and thereby diversify the investor's portfolio. The same correlation adjustment also occurs in frictionless markets, when passing from the marginal pricing measure for the pure investment problem to its counterpart in the presence of a small option position. In this sense, the marginal pricing implications of transaction costs are therefore the same as for selling a path dependent option with payoff $\int_0^T (\Delta\varphi^+_t)^2 d\langle S \rangle_t$. It is important to emphasize, however, that this is not the case at all for the corresponding hedge.

\appendix

\section{Derivation of the Main Results}\label{sec:appendix}

In the following, the main results are derived by applying formal perturbation arguments to the martingale optimality conditions for a frictionless \emph{shadow price}. The latter is a ``least favorable'' frictionless market extension in the sense that it evolves in the bid-ask spread, thereby leading to potentially more favorable trading prices, but admits an optimal policy that only entails the purchase resp.\ sale of risky shares when the shadow price coincides with the ask resp.\ bid price.  

The observation that such a shadow price should always exist can be traced back to Jouini and Kallal \cite{jouini.kallal.95} as well as Cvitani\'c and Karatzas \cite{cvitanic.karatzas.96} (also cf.\ Loewenstein \cite{loewenstein.00}). Starting with Kallsen and Muhle-Karbe \cite{kallsen.muhlekarbe.10}, this concept has recently also been used for the computation and verification of optimal policies in simple settings. Since  shadow prices are not known a priori, they have to be determined simultaneously with the optimal policy. Here, we show how to do so for general continuous asset prices, \emph{approximately} for small costs. In contrast to most previous asymptotic results, we do not first solve the problem for arbitrary costs $\varepsilon>0$ and then expand the solution around $\varepsilon=0$. Instead, we directly tackle the much simpler approximate problem for $\varepsilon \sim 0$, in the same spirit as in the approach of Soner and Touzi \cite{soner.touzi.11}.

Throughout, mathematical formalism is treated liberally. For example, we do not state and verify technical conditions warranting the uniform integrability of local martingales, interchange of integration and differentiation, and the uniformity of estimates. In particular, the Landau symbols $O(\cdot)$ and $o(\cdot)$ refer to pointwise estimates, with the implicit assumption of enough regularity in time and states to eventually turn these into an estimate of the expected utility generated by the approximately optimal policy. Rigorous proofs have been worked out in the present setting for the Black-Scholes model \cite{guasoni.muhlekarbe.11,bichuch.11}, and by Soner and Touzi \cite{soner.touzi.11} for an infinite-horizon consumption problem in a Markovian setup.

\subsection{Notation}
Throughout, we write $\phi \mal S$ for the stochastic integral $\int_0^\cdot \phi_t dS_t$. The identity process is denoted by $I_t=t$ and for any It\^o process $X$ we write $b^X$ and $\sigma^X$ for its local drift and diffusion coefficients, respectively, in the sense that $dX_t=b^X_t dt+\sigma^X_t dW_t$ for a standard Brownian motion $W$. Finally, for It\^o processes $X$ and $Y$, we denote by $c^{X,Y}_t=d\langle X,Y \rangle_t /dt$ their local quadratic covariation; if $X=Y$ we abbreviate to $d\langle X \rangle_t/dt=c_t^{X,X}=c_t^X$.

\subsection{Martingale Optimality Conditions}\label{sec:A2}

In this section, we formally derive conditions ensuring that a family $(\varphi^\varepsilon)_{\varepsilon>0}$ of frictional strategies is \emph{approximately optimal} as the spread $\varepsilon$ becomes small. For the convenience of the reader, we first briefly recapitulate their \emph{exact} counterparts in the frictionless case.

\subsubsection*{Frictionless Optimality Conditions}

In the absence of transaction costs ($\varepsilon=0$), the following duality result is well known (cf., e.g., \cite{delbaen.al.02}): The wealth process $x+\varphi \mal S$ corresponding to a trading strategy $\varphi$ is optimal, if (and essentially only if) there exists a process $Z$ satisfying the following optimality conditions:
\begin{enumerate}
\item[i)] $Z$ is a martingale.
\item[ii)] $ZS$ is a martingale.
\item[iii)] $Z_T=U'(x+\varphi \mal S_T)$.
\end{enumerate}
The first two conditions imply that $Z$ is -- up to normalization -- the density of an equivalent martingale measure $Q$ for $S$. The third identifies it as the solution to a dual minimization problem, linked to the primal maximizer by the usual first-order condition.

Let us briefly recall why conditions i)-iii) imply the optimality of $\varphi$. To this end, let $\psi$ be any competing strategy. Then, the concavity of the utility function $U$ and condition iii) imply
\begin{align*}
E[U(x+\psi \mal S_T)] &\leq E[U(x+\varphi \mal S_T)] + E[U'(x+\varphi \mal S_T)(\psi-\varphi)\mal S_T]\\
&= E[U(x+\varphi \mal S_T)] + E[Z_0] E_Q[(\psi-\varphi)\mal S_T].
\end{align*}
Since $S$ and in turn the wealth process $(\psi-\varphi)\mal S$ is a $Q$-martingale by conditions i) and ii), the second expectation vanishes and the optimality of $\varphi$ follows.

\subsubsection*{Approximate Optimality Conditions with Transaction Costs} 

Now, let us derive \emph{approximate} versions of conditions i), ii), iii) in the presence of small transaction costs $\varepsilon$, ensuring the \emph{approximate optimality} of a family $(\varphi^\varepsilon)_{\varepsilon>0}$ of strategies, at the leading order $O(\varepsilon^{2/3})$ as $\varepsilon$ becomes small. The exact optimal strategies converge to their frictionless counterpart. Hence, it suffices to consider families $(\psi^\varepsilon)_{\varepsilon>0}$ of strategies converging to the frictionless optimizer $\varphi$, i.e., $\psi^\varepsilon=\varphi+o(1)$.

Let $(\varphi^\varepsilon)_{\varepsilon>0}$ be a candidate family of strategies whose optimality we want to verify. As above, for any family of competitors $(\psi^\varepsilon)_{\varepsilon>0}$, the concavity of $U$ implies:
\begin{equation}\label{eq:concave}
E\left[U(X^{\psi^\varepsilon}_T)\right] \leq E\left[U(X^{\varphi^\varepsilon}_T)\right] + E\left[U'(X^{\varphi^\varepsilon}_T)(X^{\psi^\varepsilon}_T-X^{\varphi^\varepsilon}_T)\right],
\end{equation}
where $X_T^{\psi^\varepsilon},X_T^{\varphi^\varepsilon}$ denote the payoffs generated by trading the strategies with transaction costs. Now, suppose we can find \emph{shadow prices} $S^\varepsilon$ evolving in the bid-ask spreads $(1\pm \varepsilon)S$,  matching the trading prices $(1\pm \varepsilon)S$ in the original market with transaction costs whenever the respective strategies $\varphi^\varepsilon$ trade. Then, the frictional wealth process associated to $\varphi^\varepsilon$ evidently coincides with its frictionless counterpart for $S^\varepsilon$, i.e., $X^{\varphi^\varepsilon}_T=x+\varphi^\varepsilon \mal S^\varepsilon_T$. For any other strategy, trading in terms of $S^\varepsilon$ rather than with the original bid-ask spread can only increase wealth, since trades are carried out at potentially more favorable prices: $X^{\psi^\varepsilon}_T \leq x+\psi^\varepsilon \mal S^\varepsilon_T$. Together with \eqref{eq:concave}, this implies:
\begin{equation}\label{eq:concave2}
E\left[U(X^{\psi^\varepsilon}_T)\right] \leq E\left[U(X^{\varphi^\varepsilon}_T)\right] + E\left[U'(X^{\varphi^\varepsilon}_T)(\psi^\varepsilon-\varphi^\varepsilon)\mal S^\varepsilon_T\right].
\end{equation}
Now, suppose we can find a process $Z^\varepsilon$ satisfying the following approximate versions of i), ii), iii) above:
\begin{enumerate}
\item[$\mbox{i}^\varepsilon)$] $Z^\varepsilon$ is approximately a martingale, in that its drift rate $b^{Z^\varepsilon}$ is of order $O(\varepsilon^{2/3})$.
\item[$\mbox{ii}^\varepsilon)$] $Z^\varepsilon S^\varepsilon$ is approximately a martingale, in that its drift rate $b^{Z^\varepsilon S^{\varepsilon}}$ is of order $O(\varepsilon^{2/3})$.
\item[$\mbox{iii}^\varepsilon)$] $Z^\varepsilon_T=U'(x+\varphi^\varepsilon \mal S^\varepsilon_T)+O(\varepsilon^{2/3})$.
\end{enumerate}
Then, since $\psi^\varepsilon-\varphi^\varepsilon=o(1)$, Condition $\mbox{iii}^\varepsilon)$ implies that \eqref{eq:concave2} can be rewritten as
\begin{equation*}
E\left[U(X^{\psi^\varepsilon}_T)\right] \leq E\left[U(X^{\varphi^\varepsilon}_T)\right] + E\left[Z^\varepsilon_T ((\psi^\varepsilon-\varphi^\varepsilon)\mal S^\varepsilon_T)\right]+o(\varepsilon^{2/3}).
\end{equation*}
Applying integration by parts twice yields 
\begin{align*}
Z^\varepsilon((\psi^\varepsilon-\varphi^\varepsilon)\mal S^\varepsilon) &= Z^\varepsilon(\psi^\varepsilon-\varphi^\varepsilon)\mal S^\varepsilon+((\psi^\varepsilon-\varphi^\varepsilon)\mal S^\varepsilon)\mal Z^\varepsilon +(\psi^\varepsilon-\varphi^\varepsilon)\mal \langle Z^\varepsilon, S^\varepsilon \rangle \\
&= \left((\psi^\varepsilon-\varphi^\varepsilon)\mal S^\varepsilon-(\psi^\varepsilon-\varphi^\varepsilon)S^\varepsilon\right)\mal Z^\varepsilon +(\psi^\varepsilon-\varphi^\varepsilon)\mal (Z^\varepsilon S^\varepsilon),
\end{align*}
and in turn
\begin{equation*}
E\left[U(X^{\psi^\varepsilon}_T)\right] \leq E\left[U(X^{\varphi^\varepsilon}_T)\right] + E\left[\left((\psi^\varepsilon-\varphi^\varepsilon)\mal S^\varepsilon-(\psi^\varepsilon-\varphi^\varepsilon)S^\varepsilon\right)\mal Z_T^\varepsilon +(\psi^\varepsilon-\varphi^\varepsilon)\mal (Z^\varepsilon S^\varepsilon)_T\right]+o(\varepsilon^{2/3}).
\end{equation*}
The second expectation is given by the integrated expected drift rate of its argument,
$$
\left((\psi^\varepsilon-\varphi^\varepsilon)\mal S^\varepsilon-(\psi^\varepsilon-\varphi^\varepsilon)S^\varepsilon\right)b^{Z^{\varepsilon}}+(\psi^\varepsilon-\varphi^\varepsilon)b^{Z^\varepsilon S^\varepsilon},$$
which is of order $o(\varepsilon^{2/3})$, by conditions $\mbox{i}^\varepsilon)$ and $\mbox{ii}^\varepsilon)$ above and because $\psi^\varepsilon-\varphi^\varepsilon=o(1)$. Hence,
$$
E\left[U(X^{\psi^\varepsilon}_T)\right] \leq E\left[U(X^{\varphi^\varepsilon}_T)\right] + o(\varepsilon^{2/3}),
$$
and the expected utilities of the candidate family $(\varphi^\varepsilon)_{\varepsilon>0}$ therefore dominate those of the competitors $(\psi^\varepsilon)_{\varepsilon>0}$ at the leading order $O(\varepsilon^{2/3})$. In summary, the strategies $(\varphi^\varepsilon)_{\varepsilon>0}$ are indeed approximately optimal if we can find a shadow price $S^\varepsilon$ and an approximate martingale density $Z^\varepsilon$ satisfying the approximate optimality conditions $\mbox{i}^\varepsilon)$, $\mbox{ii}^\varepsilon)$, $\mbox{iii}^\varepsilon)$.

\subsection{Derivation of a Candidate Policy}

We now look for strategies $\varphi^\varepsilon$, shadow prices $S^\varepsilon$, and approximate martingale densities $Z^\varepsilon$ satisfying the approximate optimality conditions $\mbox{i}^\varepsilon)-\mbox{iii}^\varepsilon)$. Write
$$\varphi^\varepsilon=\varphi+\Delta\varphi, \quad S^\varepsilon=S+\Delta S.$$
Motivated by previous asymptotic results \cite{whalley.wilmott.97,guasoni.muhlekarbe.11,bichuch.11,soner.touzi.11}, we assume that the deviations of the optimal strategy with transaction costs from the frictionless optimizer are asymptotically proportional to the cubic root of the spread:
\begin{equation}\label{eq:deltavarphi}
\Delta\varphi=O(\varepsilon^{1/3}).
\end{equation}
Since the shadow price $S^\varepsilon$ has to lie in the bid-ask spread $(1\pm \varepsilon)S$, we must have 
\begin{equation}\label{eq:deltas}
\Delta S=O(\varepsilon).
\end{equation} 
In addition, we assume that $\Delta S$ is an It\^o process with drift and diffusion coefficients satisfying:
\begin{equation}\label{eq:dynamicsdeltas}
b^{\Delta S}=O(\varepsilon^{1/3}), \quad \sigma^{\Delta S}=O(\varepsilon^{2/3}).
\end{equation}
All of these assumptions will turn out to be consistent with the results of our calculations below, see Section \ref{sec:approxopt}. Now, notice that 
$$\varphi^\varepsilon \mal S^\varepsilon =\varphi \mal S +\Delta\varphi \mal S +O(\varepsilon^{2/3}),$$
because \eqref{eq:deltavarphi} and \eqref{eq:dynamicsdeltas} give $\Delta\varphi \mal \Delta S=O(\varepsilon^{2/3})$, and integration by parts in conjunction with \eqref{eq:deltas} and \eqref{eq:dynamicsdeltas} shows $\varphi \mal \Delta S = O(\varepsilon^{2/3})$. Therefore, 
$$ U'(x+\varphi^\varepsilon \mal S^\varepsilon_T) =p e^{-p(x+\varphi^\varepsilon \mal S^\varepsilon_T )} =p e^{-p(x+\varphi \mal S_T)}(1-p\Delta\varphi \mal S_T) +O(\varepsilon^{2/3}).$$
The factor $p e^{-p(x+\varphi \mal S_T)}$ coincides with the terminal value of the frictionless martingale density $Z$ (cf.\ the frictionless optimality condition iii) above). The process
$$Z^\varepsilon= Z(1-p \Delta \varphi \mal S)$$
therefore is a martingale (because $Z$ is the density of a martingale measure for $S$), satisfying Condition $\mbox{i}^\varepsilon)$, for which $\mbox{iii}^\varepsilon)$ holds as well. It remains to determine $\Delta S$ and $\Delta\varphi$ for which $\mbox{ii}^\varepsilon)$ holds, too. Integration by parts yields
$$Z^\varepsilon S^\varepsilon-Z^\varepsilon_0 S^\varepsilon_0=S^\varepsilon\mal Z^\varepsilon+Z^\varepsilon\mal S^\varepsilon+\langle Z^\varepsilon,S^\varepsilon \rangle.$$
Since the martingale $Z^\varepsilon$ has zero drift, it follows that the drift rate of $Z^\varepsilon S^\varepsilon$ is given by
\begin{equation}\label{eq:bZS}
b^{Z^\varepsilon S^\varepsilon}=Z^\varepsilon(b^S+b^{\Delta S})+c^{Z^\varepsilon,S+\Delta S}.
\end{equation}
As $b^{\Delta S}=O(\varepsilon^{1/3})$ by assumption, and $Z^\varepsilon=Z(1-p\Delta\varphi \mal S)$, it follows that
\begin{equation}\label{eq:drift}
Z^\varepsilon(b^S+b^{\Delta S})=Z\left(b^S-p(\Delta\varphi\mal S)b^S+b^{\Delta S}\right)+O(\varepsilon^{2/3}).
\end{equation}
Moreover, writing the frictionless martingale density as a stochastic exponential $Z=\scr{E}(N)=1+Z \mal N$, it follows from \eqref{eq:dynamicsdeltas} and integration by parts that
\begin{align*}
\langle Z^\varepsilon,S+\Delta S\rangle = & \langle Z(1-p\Delta\varphi \mal S),S \rangle+O(\varepsilon^{2/3})\\
=& Z \mal \left( \langle N, S \rangle -p(\Delta \varphi \mal S) \mal \langle N,S \rangle  - p \Delta \varphi \mal  \langle S,S \rangle\right)+O(\varepsilon^{2/3}),
\end{align*}
so that 
\begin{equation}\label{eq:cZS}
c^{Z^\varepsilon,S+\Delta S}=Z\left(c^{N,S}-p(\Delta\varphi\mal S)c^{N,S}-p\Delta\varphi c^{S}\right)+O(\varepsilon^{2/3}).
\end{equation}
Then, inserting \eqref{eq:drift} and \eqref{eq:cZS} into \eqref{eq:bZS} and using that $b^S+c^{N,S}=0$ by Girsanov's theorem because $ZS=\scr{E}(N)S$ is a martingale by the frictionless optimality condition ii), gives
\begin{align*}
b^{Z^\varepsilon S^\varepsilon}&=Z\left(b^S+c^{N,S}-p(\Delta\varphi\mal S)(b^S+c^{N,S})+b^{\Delta S}-p\Delta\varphi c^{S}\right)+O(\varepsilon^{2/3})\\
&=Z(b^{\Delta S}-p\Delta\varphi c^{S})+O(\varepsilon^{2/3}).
\end{align*}
To make the drift of $Z^\varepsilon S^\varepsilon$ vanish -- up to terms of order $O(\varepsilon^{2/3})$ -- in accordance with $\mbox{ii}^\varepsilon)$, it is therefore necessary that 
\begin{equation}\label{eq:bC}
b^{\Delta S}=p\Delta\varphi c^{S}+O(\varepsilon^{2/3}).
\end{equation}
This drift condition naturally leads to an ansatz of the form $\Delta S=f(\Delta \varphi)$. Then, since the shadow price $S^\varepsilon=S+\Delta S$ has to move from the ask price $(1+\varepsilon)S$ to the bid price $(1-\varepsilon)S$ as $\Delta\varphi$ varies between some buy boundary $\Delta\varphi^-$ and some sell boundary $\Delta\varphi^+$, the function $f$ has to satisfy the boundary conditions
\begin{equation}\label{eq:boundary}
f(\Delta\varphi^-)=\varepsilon S, \quad f(\Delta\varphi^+)=-\varepsilon S.
\end{equation}
Moreover, even though the process $\Delta\varphi$ is reflected to remain between the trading boundaries, these singular terms should vanish in the dynamics of $\Delta S$, so that the shadow price $S^\varepsilon=S+\Delta S$ does not allow for arbitrage. By It\^o's formula, this implies that the derivative of $f$ should vanish at the boundaries:
\begin{equation}\label{eq:pasting}
f'(\Delta\varphi^-)=0, \quad f'(\Delta\varphi^+)=0.
\end{equation}
The simplest family of functions capable of matching these boundary conditions is given by the symmetric cubic polynomials
$$f(x)=\alpha x^3- \gamma x.$$
With this ansatz, \eqref{eq:pasting} gives
$$\Delta\varphi^\pm =\pm \sqrt{\frac{\gamma}{3\alpha}},$$
and \eqref{eq:boundary} implies
$$\gamma=\left(\frac{1}{2}\varepsilon S\right)^{2/3}3\alpha^{1/3}.$$
Moreover, It\^o's formula applied to $f(x)=\alpha x^3-\gamma x$ yields
$$\Delta S-\Delta S_0=(3 \alpha \Delta\varphi^2-\gamma)\mal \Delta\varphi+ 3\alpha \Delta\varphi \mal \langle \Delta\varphi, \Delta \varphi \rangle.$$
Now, notice that the optimal trading strategy $\varphi^\varepsilon=\varphi+\Delta\varphi$ with transaction costs is necessarily of finite variation. Assuming it is also continuous then implies $\langle \Delta\varphi \rangle=\langle \varphi \rangle$. Moreover, since $\varphi^\varepsilon$ is constant except at the trading boundaries (where $\Delta \varphi=\Delta\varphi^\pm$ and in turn $3 \alpha \Delta\varphi^2-\gamma=0$), we also have 
$$(3 \alpha \Delta\varphi^2-\gamma)\mal \Delta\varphi=-(3 \alpha\Delta\varphi^2-\gamma)\mal\varphi.$$
Thus,
$$\Delta S-\Delta S_0=-(3 \alpha\Delta\varphi^2-\gamma)\mal\varphi+ 3\alpha\Delta\varphi \mal \langle \varphi \rangle,$$
and the drift coefficient of $\Delta S$ is given by
$$b^{\Delta S}=3\alpha\Delta\varphi c^{\varphi}+O(\varepsilon^{2/3}).$$
Comparing this to the leading-order term in \eqref{eq:bC}, we obtain
$$\alpha=\frac{p}{3} \frac{c^{S}}{c^{\varphi}},$$
and in turn
$$\gamma=\left(\frac{3p^{1/2}}{2}\sqrt{\frac{c^{S}}{c^{\varphi}}}S\right)^{2/3}\varepsilon^{2/3}$$
as well as
$$\Delta\varphi^{\pm}=\pm \sqrt{\frac{\gamma}{3\alpha}}=\pm\left(\frac{3}{2p}\frac{c^{\varphi}}{c^{S}}\varepsilon S\right)^{1/3}.$$
At the first order, this determines the optimal strategy $\varphi^\varepsilon$ with transaction costs as the minimal amount of trading necessary to remain in the randomly changing interval $[\varphi+\Delta\varphi^-,\varphi+\Delta\varphi^+]$ around the frictionless optimizer $\varphi$.

\subsection{Approximate Optimality}\label{sec:approxopt}
The above considerations assumed that the coefficients $\alpha, \gamma$ are constant, but then lead to stochastic processes $\alpha_t,\gamma_t$, which seems contradictory at first glance. However, we can verify a fortiori that this choice does indeed satisfy $\mbox{i}^\varepsilon)-\mbox{iii}^\varepsilon)$. To see this set, for $\alpha,\gamma$ as above,
$$\Delta S_t=\alpha_t \Delta\varphi_t^3-\gamma_t \Delta\varphi_t$$
and let the strategy $\varphi^\varepsilon=\varphi+\Delta\varphi$ correspond to the minimal amount of trading necessary to remain within the boundaries $\Delta\varphi^{\pm}$ around the frictionless optimizer $\varphi$. Then by definition, the process $S^\varepsilon:=S+\Delta S$ takes values in the bid-ask spread $[(1-\varepsilon)S,(1+\varepsilon)S]$ and coincides with the bid resp.\ ask price whenever $\varphi^\varepsilon$ reaches the selling boundary $\varphi+\Delta\varphi^+$ resp.\ the buying boundary $\varphi-\Delta\varphi^+$ as required for a shadow price. Concerning the dynamics of $\Delta S$, notice that integration by parts (now taking into account the stochasticity of $\alpha$ and $\gamma$) and It\^o's formula give
\begin{align}
\Delta S-\Delta S_0 =&\alpha\mal (\Delta\varphi^3) +\Delta\varphi^3 \mal \alpha+\langle \alpha,\Delta\varphi^3\rangle -\gamma \mal \Delta\varphi- \Delta\varphi \mal \gamma - \langle \gamma,\Delta\varphi \rangle \notag \\
=&(3 \alpha \Delta\varphi^2-\gamma)\mal \Delta \varphi+(3\alpha\Delta\varphi)\mal \langle \Delta\varphi\rangle \notag \\
&\quad+\Delta\varphi^3 \mal \alpha -\Delta\varphi \mal \gamma +(3\Delta\varphi^2) \mal \langle \alpha, \Delta\varphi \rangle -\langle \gamma,\Delta\varphi \rangle \notag \\
=&-(3 \alpha \Delta\varphi^2-\gamma)\mal \varphi+(3\alpha\Delta\varphi)\mal \langle \varphi\rangle \notag \\
&\quad+\Delta\varphi^3 \mal \alpha -\Delta\varphi \mal \gamma -(3\Delta\varphi^2) \mal \langle \alpha,\varphi \rangle +\langle \gamma,\varphi \rangle.\label{eq:dds}
\end{align}
Here we have used for the last equality that $\varphi^{\varepsilon}=\varphi+\Delta\varphi$ only moves on the set $\Delta\varphi=\Delta\varphi^\pm$ where $3 \alpha \Delta\varphi^2-\gamma=0$, and that $\Delta\varphi=-\varphi +\varphi^\varepsilon$ only differs from $-\varphi$ by a finite variation term. If the risky asset $S$, the frictionless optimizer $\varphi$, as well as their local quadratic variation processes $c^{S}, c^{\varphi}$ (and in turn the processes $\alpha$ and $\gamma$) follow sufficiently regular It\^o processes, this representation shows that this property is passed on to $\Delta S$. Moreover, since $\Delta\varphi=O(\varepsilon^{1/3})$, $\gamma=O(\varepsilon^{2/3})$, and $\alpha=O(1)$ (and the same asymptotics are valid for the drift and diffusion coefficients of $\alpha$ and $\gamma$), its diffusion coefficient is indeed of order $O(\varepsilon^{2/3})$ and its drift rate is of order $O(\varepsilon^{1/3})$. More specifically, the latter is given by $b^{\Delta S}=3\alpha \Delta\varphi c^{\varphi}+O(\varepsilon^{2/3})$; hence, by definition of $\alpha$, the drift condition \eqref{eq:bC} and in turn the approximate optimality condition $\mbox{ii}^{\varepsilon})$ is indeed satisfied for the shadow price $S^\varepsilon$ and the strategy $\varphi^\varepsilon$. Consequently, the latter is approximately optimal for small spreads.

\subsection{Computation of the Leading-Order Utility Loss}

Let us now compute -- at the leading order $O(\varepsilon^{2/3})$ -- the expected utility that can be obtained by applying the strategy $\varphi^\varepsilon$. Since the latter is approximately optimal, this will then also determine the leading-order impact of transaction costs on the certainty equivalent of trading optimally in the market.

To do this, the analysis of the previous section needs to be refined. Including a second term in the Taylor expansion of the utility function, and taking into account $\varphi^\varepsilon \mal S^\varepsilon =\varphi \mal S+\Delta\varphi \mal S+\varphi^\varepsilon \mal \Delta S$, where $\varphi^\varepsilon \mal \Delta S$ is of order $O(\varepsilon^{2/3})$,\footnote{This follows using integration by parts to write $\varphi^\varepsilon \mal \Delta S=\Delta\varphi \mal \Delta S + \varphi \Delta S -\varphi_0 \Delta S_0 -\Delta S \mal \varphi -\langle \varphi, \Delta S \rangle$, and recalling that the drift and diffusion coefficients of $\Delta S$ are of order $O(\varepsilon^{1/3})$ and $O(\varepsilon^{2/3})$, respectively, whereas $\Delta S$ and $\Delta \varphi$ are of order $O(\varepsilon)$ resp.\ $O(\varepsilon^{1/3})$.} gives:
\begin{align*}
E[U(x+\varphi^\varepsilon \mal S^\varepsilon_T)] = E[U(x+\varphi \mal S_T)]&+E[U'(x+\varphi \mal S_T)(\Delta \varphi \mal S_T +\varphi^\varepsilon \mal \Delta S_T)]\\
&+\frac{1}{2} E[U''(x+\varphi \mal S_T)(\Delta\varphi \mal S_T)^2]+O(\varepsilon).
\end{align*}
For the exponential utility function $U(x)=-e^{-px}$, the marginal utility $U'(x+\varphi \mal S_T)$ needs to be normalized by $E[U'(x+\varphi \mal S_T)]=-p E[U(x+\varphi\mal S_T)]$ to obtain the density of an equivalent martingale measure $Q$ for $S$. Since, moreover, the absolute risk-aversion $-U''/U'=p$ is constant, it follows that
\begin{align*}
E[U(x+\varphi^\varepsilon \mal S^\varepsilon_T)] = E[U(x+\varphi \mal S_T)]\left(1-pE_Q[\varphi^\varepsilon \mal \Delta S_T]+\frac{p^2}{2} E_Q[(\Delta\varphi \mal S_T)^2]\right)+O(\varepsilon),
\end{align*}
where we have used that the expectation of the $Q$-martingale $\Delta \varphi \mal S$ vanishes. The second correction term $\frac{p^2}{2} E_Q[(\Delta\varphi \mal S_T)^2]$ represents the leading-order relative utility loss due to displacement, incurred by trading $\varphi^\varepsilon$ instead of the frictionless optimizer $\varphi$ at the frictionless price $S$. The first correction term $-pE_Q[\varphi^\varepsilon \mal \Delta S_T]$ measures the utility loss incurred directly due to transaction costs, when trades are carried out at the shadow price $S^\varepsilon$ rather than at the mid price $S$.

Let us first focus on the displacement loss $\frac{p^2}{2} E_Q[(\Delta\varphi \mal S_T)^2]$. Integration by parts gives
$$(\Delta\varphi \mal S_T)^2=2(\Delta\varphi \mal S)\Delta\varphi \mal S_T +\Delta \varphi^2 \mal \langle S \rangle_T.$$
As the first term is a $Q$-martingale, it follows that the leading-order displacement loss is given by
$$\frac{p^2}{2} E_Q[(\Delta\varphi \mal S_T)^2]=\frac{p^2}{2} E_Q[\Delta\varphi^2 \mal \langle S \rangle_T].$$
Now, consider the direct transaction cost loss $-pE_Q[\varphi^\varepsilon \mal \Delta S_T]$. Integration by parts and $\Delta S=O(\varepsilon)$ yield
$$\varphi \mal \Delta S= -\langle \varphi, \Delta S \rangle +O(\varepsilon).$$
First taking into account the dynamics of $\Delta S$ (cf.\ \eqref{eq:dds}), and then inserting the definitions of $\alpha$, $\gamma$ and the trading boundaries $\Delta\varphi^+$, as well as $c^{S} \mal I=\langle S \rangle$, it follows that
\begin{align*}
-p E_Q[\varphi \mal \Delta S_T]&=-p E_Q[(3\alpha \Delta\varphi^2 -\gamma)c^{\varphi} \mal I_T]+O(\varepsilon)=-p^2E_Q[(\Delta \varphi^2-(\Delta\varphi^+)^2)\mal \langle S \rangle_T]+O(\varepsilon).
\end{align*}
The remaining term $-pE_Q[\Delta\varphi \mal \Delta S_T]$ can again by computed by integrating the drift rate of the argument of the expectation (here, $b^{\Delta S,Q}$ denotes the drift of $\Delta S$ under the measure $Q$):
\begin{align*}
-p E_Q[\Delta\varphi \mal \Delta S_T] &= -p E_Q[\Delta \varphi b^{\Delta S,Q} \mal I_T]\\
&=-p E_Q[\Delta\varphi (b^{\Delta S}+c^{N,\Delta S})\mal I_T]=-p E_Q[\Delta\varphi b^{\Delta S} \mal I_T]+O(\varepsilon).
\end{align*}
Here, the second equality follows from Girsanov's theorem, and the third one holds since $\Delta\varphi=O(\varepsilon^{1/3})$ and the diffusion coefficient of $\Delta S$ is of order $O(\varepsilon^{2/3})$  by \eqref{eq:dds}. Combined with the drift condition \eqref{eq:bC} and $c^{S} \mal I=\langle S\rangle$, this yields
$$-p E_Q[\Delta\varphi \mal \Delta S_T]=-p^2 E_Q[\Delta\varphi^2 \mal \langle S \rangle_T]+O(\varepsilon).$$
As a consequence, the total relative utility loss directly caused by transaction costs is given by
$$-pE_Q[\varphi^\varepsilon \mal \Delta S_T]=p^2 E_Q[((\Delta\varphi^+)^2-2\Delta\varphi^2)\mal \langle S \rangle_T]+O(\varepsilon).$$
To further simplify the formulas for both parts of the utility loss, replace -- at the leading order $O(\varepsilon^{2/3})$ -- the terms $\Delta\varphi^2$ by their expectation $\frac{1}{3}(\Delta\varphi^+)^2$ under the uniform distribution on $[\Delta\varphi^-,\Delta\varphi^+]$ (compare \cite{rogers.04,goodman.ostrov.10}), which is justified below. Then, the displacement loss is determined as $\frac{p^2}{6} E_Q[(\Delta\varphi^+)^2 \mal \langle S,S \rangle_T]+o(\varepsilon^{2/3})$, and the transaction cost loss is found to be given by twice that value. Hence, the total utility loss due to transaction costs is given by
$$E[U(x+\varphi^\varepsilon \mal S^\varepsilon_T)] = E[U(x+\varphi \mal S_T)]\left(1+ \frac{p^2}{2} E_Q\left[(\Delta\varphi^+)^2 \mal \langle S \rangle_T\right]\right)+o(\varepsilon^{2/3}),$$
and the claimed formula for the certainty equivalent follows by taking logarithms and Taylor expansion.

To complete the argument, it remains to verify that we can indeed pass to the uniform distribution for $\Delta\varphi$ at the leading order. To this end, define $D=\Delta\varphi \sigma^S$, which is an It\^o process reflected to stay between the boundaries $D^{\pm}=\Delta\varphi^\pm \sigma^S$. In the interior of $[D^-,D^+]$, the strategy $\varphi^\varepsilon$ is constant, so that $\Delta\varphi=-\varphi$. Hence, the drift rate $b^D$ and the diffusion coefficient $\sigma^D$ of $D$ are both of order $O(1)$. Now, fix a mesh $0=t^\varepsilon_0<\ldots<t^\varepsilon_{N^\varepsilon}=T$ with mesh size of order $O(\varepsilon^{1/3})$, and write
\begin{equation}\label{eq:mesh} 
\int_0^T \Delta\varphi_u^2 d\langle S \rangle_u= \int_0^T D_u^2 du= \sum_{i=1}^{N^\varepsilon} \int_{t^\varepsilon_{i-1}}^{t^\varepsilon_{i}} D_u^2 du.
\end{equation}
Rescale $D$ by dividing by $\varepsilon^{1/3}$ and integrating over $v=u/\varepsilon^{2/3}$ instead of $u$, obtaining 
\begin{equation}\label{eq:rescaled}
\int_{t^\varepsilon_{i-1}}^{t^\varepsilon_{i}} D_u^2 du= \varepsilon^{4/3} \int_{t^\varepsilon_{i-1}/\varepsilon^{2/3}}^{t^\varepsilon_{i}/\varepsilon^{2/3}}\left(\frac{D_{\varepsilon^{2/3}v}}{\varepsilon^{1/3}}\right)^2 dv.
\end{equation}
The drift and diffusion coefficient of the rescaled integrand $(\varepsilon^{-1/3} D_{\varepsilon^{2/3} v})_{v \geq 0}$ are given by $b^D_{\varepsilon^{2/3}v}\varepsilon^{1/3}=O(\varepsilon^{1/3})$ and $\sigma^D_{\varepsilon^{2/3}v}=O(1)$, respectively. Hence, at the leading order, it equals driftless Brownian motion with constant volatility $\sigma^D_{t^\varepsilon_i}$ on $t^\varepsilon_{i-1}/\varepsilon^{2/3} \leq v \leq t^\varepsilon_{i}/\varepsilon^{2/3}$. Reflected Brownian motion on the interval $[D^-_{\varepsilon^{2/3}v}/\varepsilon^{1/3},D^+_{\varepsilon^{2/3}v}/\varepsilon^{1/3}]$ (whose boundaries are constant and equal to $D^\pm_{t^\varepsilon_{i-1}}$, at the leading order) has a uniform stationary distribution with second moment $(D_{t^\varepsilon_{i-1}}^+)^2/3\varepsilon^{2/3}$. Consequently, the ergodic theorem \cite[II.35 and II.36]{borodin.salminen.02} implies
$$\int_{t^\varepsilon_{i-1}/\varepsilon^{2/3}}^{t^\varepsilon_{i}/\varepsilon^{2/3}}\left(\frac{D_{\varepsilon^{2/3}v}}{\varepsilon^{1/3}}\right)^2 dv = \frac{t^\varepsilon_{i}-t^\varepsilon_{i-1}}{\varepsilon^{2/3}} \left(\frac{(D^+_{t^\varepsilon_{i-1}})^2}{3\varepsilon^{2/3}}+o(1)\right).$$
Combining this with \eqref{eq:mesh} and \eqref{eq:rescaled} and letting the mesh size go to zero, we obtain the assertion:
$$\int_0^T \Delta\varphi_u^2 d\langle S,S\rangle_u= \int_0^T D_u^2 du = \frac{1}{3} \int_0^T (D_u^+)^2 du+o(\varepsilon^{2/3}).$$

\bibliographystyle{abbrv}
\bibliography{tractrans}

\end{document}